\documentclass[letter]{jpsj2} 
\title{
A Novel Pyrochlore Ruthenate: Ca$_{2}$Ru$_{2}$O$_{7}$
}

\author{Taiya \textsc{Munenaka} and Hirohiko \textsc{Sato}\thanks{E-mail address: hirohiko@phys.chuo-u.ac.jp}
}

\inst{Department of Physics, Chuo University, 1-13-27 Kasuga, Bunkyo-ku,
Tokyo 112-8551, Japan
 \\
}

\abst{Single crystals of a novel ruthenate, Ca$_{2}$Ru$_{2}$O$_{7}$, were obtained. An X-ray diffraction study on a single crystal revealed that this material crystallizes in a pyrochlore structure with a lattice parameter, $a = 10.197$ \AA. The magnetic susceptibility above 30 K is the summation of a Curie-Weiss contribution and a constant term independent of temperature. The effective moment per Ru atom is only 0.36 $\mu_{\mbox{\scriptsize B}}$, one order of magnitude smaller than that expected from a localized spin model with $S=3/2$ for Ru$^{5+}$. Below 23 K, the localized spins freeze in a spin-glass state. The resistivity at room temperature is $2 \times 10^{-3}$ $\Omega$cm, comparable to that in metallic, highly correlated oxides.}

\kword{ruthenate, pyrochlore, magnetism, electric conduction, spin glass
}

\begin{document}
\maketitle


Frustration results in many types of unexpected phenomena. Among three-dimensional frustrated systems, the pyrochlore lattice is particularly interesting for its strong frustration originating from a purely geometric reason. A typcial pyrochlore oxide has the composition A$_{2}$B$_{2}$O$_{7}$. In this system, the B sites (and also the A sites) form a three-dimensional network based on the B$_{4}$ tetrahedron. From another point of view, we can regard the pyrochlore lattice as a three-dimensional version of a Kagom\'e lattice. Therefore, perfect geometric frustration is inherent in this structure, and many interesting phenomena emerge. For example, Y$_{2}$Mo$_{2}$O$_{7}$ exhibits spin-glass behavior,\cite{gingras97,gardner99,miyoshi00} demonstrating that the geometric frustration due to the antiferromagnetic pyrochlore lattice itself is responsible for the glassy state, even if there is no structural disorder. On the other hand, the nearest-neighbor interaction is ferromagnetic in Ho$_{2}$Ti$_{2}$O$_{7}$ and Dy$_{2}$Ti$_{2}$O$_{7}$. In this case, single-ion magnetic anisotropy causes another type of frustration and consequently, ``spin ice'' behavior appears.\cite{harris97,ramirez99,higashinaka05}

Conductive pyrochlores are also remarkable systems. For Nd$_{2}$Mo$_{2}$O$_{7}$, there was the epoch-making interpretation that the anomalous Hall effect detects the chirality of Nd magnetic moments.\cite{taguchi03} A theoretical study proved that the Berry phase plays an important role in systems with a chiral spin arrangement.\cite{onoda03} Tl$_{2}$Mn$_{2}$O$_{7}$ has metallic conductivity and undergoes a ferromagnetic transition. Near the transition temperature, a giant magnetoresistance appears\cite{shimakawa96}. In Cd$_{2}$Re$_{2}$O$_{7}$, a superconducting transition was discovered at 1.5 K.\cite{sakai01,hanawa01} Furthermore, $\beta$-type pyrochlore osmates, AOs$_{2}$O$_{6}$ (A = K, Rb, Cs), also exhibit superconductivity with a relatively high $T_{c}$ (9.7 K for A = K).\cite{yonezawa04,yonezawa04b,hiroi04,hiroi05}

While searching for new materials with exotic electronic states, we have become interested in pyrochlore ruthenates. Because ruthenium 4$d$-orbitals have a character intermediate between localized and itinerant orbitals, a variety of electronic phases appear. In particular, the discovery of spin-triplet superconductivity in Sr$_{2}$RuO$_{4}$\cite{maeno94,ishida98} has aroused the interest of many material scientists.

Ruthenates with pyrochlore structures have also been actively investigated. Bi$_{2}$Ru$_{2}$O$_{7}$ and Pb$_{2}$Ru$_{2}$O$_{6.5}$ are metallic with Pauli paramagnetism,\cite{longo69,cox83,hsu88} whereas Ln$_{2}$Ru$_{2}$O$_{7}$ and Y$_{2}$Ru$_{2}$O$_{7}$ are insulators with localized magnetic moments.\cite{aleonard62,subramanian83,yoshii99,ito00} Tl$_{2}$Ru$_{2}$O$_{7}$ undergoes a metal-insulator transition.\cite{takeda98} These observations reveal that pyrochlore ruthenates display a variety of electronic phases, widely distributed over the Mott boundary. Their electronic states are very sensitive to the Ru-O distance or the Ru-O-Ru bond angle, which is related to the radius of the cations on the A site. In addition to controlling the band width by changing the cation radius, filling control of the $4d$-band also seems important in searching for novel electronic phases. However, there have been few trials\cite{yoshii99} on controlling the band filling of pyrochlore ruthenates. This is probably because the cation on the A site is trivalent in most stable pyrochlore ruthenates. Apart from Cd$_{2}$Ru$_{2}$O$_{7}$\cite{wang98}, there are no reports on stoichiometric pyrochlores composed of only Ru$^{5+}$. In the present study, we succeeded in synthesizing a new pyrochlore ruthenate with Ru$^{5+}$, Ca$_{2}$Ru$_{2}$O$_{7}$, by maintaining a high-oxidization atmosphere.

  Single crystals of Ca$_{2}$Ru$_{2}$O$_{7}$ were synthesized by a hydrothermal method. A mixture of RuO$_{2}$ (40 mg), obtained by oxidizing Ru metal (Furuya Metals, 99.99\% purity), CaO (34 mg, Soekawa Chemical, 99.99\% purity), and 0.3 ml of 30\% H$_{2}$O$_{2}$ solution was encapsulated in a gold tube. Then, it was kept in an autoclave  under 150 MPa hydrostatic pressure at 600$^{\circ}$C for 3 days. The chemical composition was determined using an energy dispersive X-ray spectrometer (EDS) installed on a scanning electron microscope. The crystal structure was analyzed using a single crystal and an imaging-plate X-ray diffractometer (Rigaku, R-Axis RAPID), in which Mo-K$\alpha$ radiation was generated using an X-ray tube and monochromized using graphite. We also used a powder X-ray diffractometer to check whether there was any contamination due to impurity phases. The magnetic susceptibility between 2 and 400 K was measured using a superconducting quantum-interference-device magnetometer. In the measurement, approximately 10 mg of nonoriented single crystals were wrapped in a piece of aluminum foil. The resistivity was measured by a DC four-wire method on an array of single crystals, connected with each other, in a closed-cycle helium refrigerator whose temperature range was between 5.5 and 300 K. The array was composed of four single crystals, and we attached the voltage leads to the same crystal located at the center. Therefore, we consider that the observed resistivity approximately reflects the behavior of a single crystal.

\begin{figure}[tb]
\begin{center}
\includegraphics[width=1.0\linewidth]{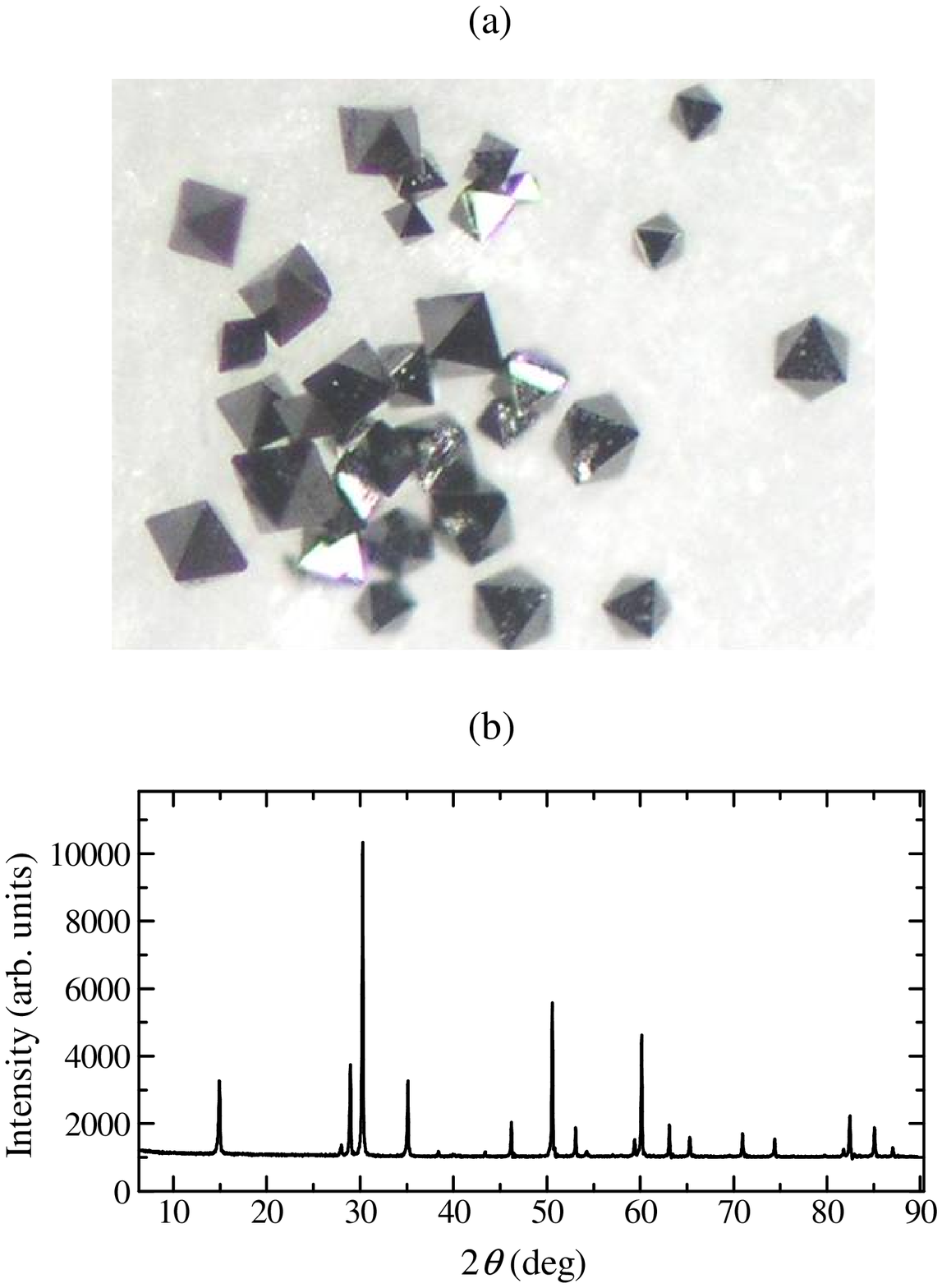}
\end{center}
\caption{(a) Micrograph of single crystals of Ca$_{2}$Ru$_{2}$O$_{7}$. The typical dimensions of the crystals are $0.1 \times 0.1 \times 0.1$ mm$^{3}$. (b) X-ray powder pattern of batch used for magnetic measurement. The whole batch was ground before measurement. The weak peaks at 28.1$^{\circ}$ and at 54.3$^{\circ}$ are from traces of RuO$_{2}$.}
\label{fig1}
\end{figure}

The obtained materials were black crystals with an octahedral shape. Observations using a microscope did not detect any other type of crystal as shown in Fig.~\ref{fig1}(a). An EDS analysis showed that the atomic compositional ratio of Ca and Ru is almost 1:1. The single-crystal X-ray diffraction revealed an F-type cubic unit cell with $a = 10.197$ \AA, which is very close to 10.143 \AA{} for Y$_{2}$Ru$_{2}$O$_{7}$\cite{kennedy95}. This strongly suggested that the our material has a pyrochlore structure. A further structural refinement was carried out by observing about 2000 reflections at room temperature. The results are summarized in Table~\ref{table1},
\begin{table}[tb]
\caption{Fractional atomic coordinates and equivalent isotropic displacement parameters 
 (\AA$^{2}$) for Ca$_{2}$Ru$_{2}$O$_{7}$. The lattice symmetry and the space group 
are \textit{cubic} and $Fd\bar{3}m$ (\#227), respectively. The lattice parameters are  $a = 10.197(2)$ \AA{}, 
$V = 1060.4(3)$ \AA{}$^3$ and $Z=8$. The final reliability factor is $R(F) = 2.7 \%$ for 
1888 observed reflections.}
\label{table1}
\begin{tabular}{cccccc}
\hline
Atom & Position & $x$ & $y$ & $z$ & $B_{eq}$\\
\cline{1-6}
Ru(1) & 16$c$ & 0 &  0 &  0 &  0.602(4) \\
Ca(1) & 16$d$ & 0.5 &  0.5    &  0.5    &  2.587(9) \\
O(1)  & 48$f$ & 0.3219(1)  &  0.125    &  0.125    &  1.48(2) \\
O(2)  & 8$b$ & 0.375  &  0.375  &  0.375  &  3.65(3) \\
\hline
\end{tabular}
\end{table}
and they coincide with those for a pyrochlore structure. The analysis did not detect any clear evidence that the composition deviates from the ideal pyrochlore, although the temperature factors of Ca(1) and O(2) seem unexpectedly large. We also analyzed the powder X-ray diffraction pattern. As shown in Fig.~\ref{fig1}(b), almost all of the peaks are in agreement with those of a pyrochlore lattice, except for two weak peaks from the traces of RuO$_{2}$.
To our best knowledge, there have been no reports on the existence of the Ca$_{2}$Ru$_{2}$O$_{7}$ phase despite many studies on pyrochlore ruthenates. There has only been a study on the mixed crystal Y$_{2-x}$Ca$_{x}$$_{2}$Ru$_{2}$O$_{7}$,\cite{yoshii99} although the upper limit of $x$ was 0.6. The reason why Ca$_{2}$Ru$_{2}$O$_{7}$ has not been successfully obtained is probably that a strong oxidization atmosphere is necessary for maintaining the Ru$^{5+}$ valence at high temperatures. We suppose that a hydrothermal reaction using a strong oxidant, H$_{2}$O$_{2}$, is advantageous for realizing this condition.

\begin{figure}[tb]
\begin{center}
\includegraphics[width=0.8\linewidth]{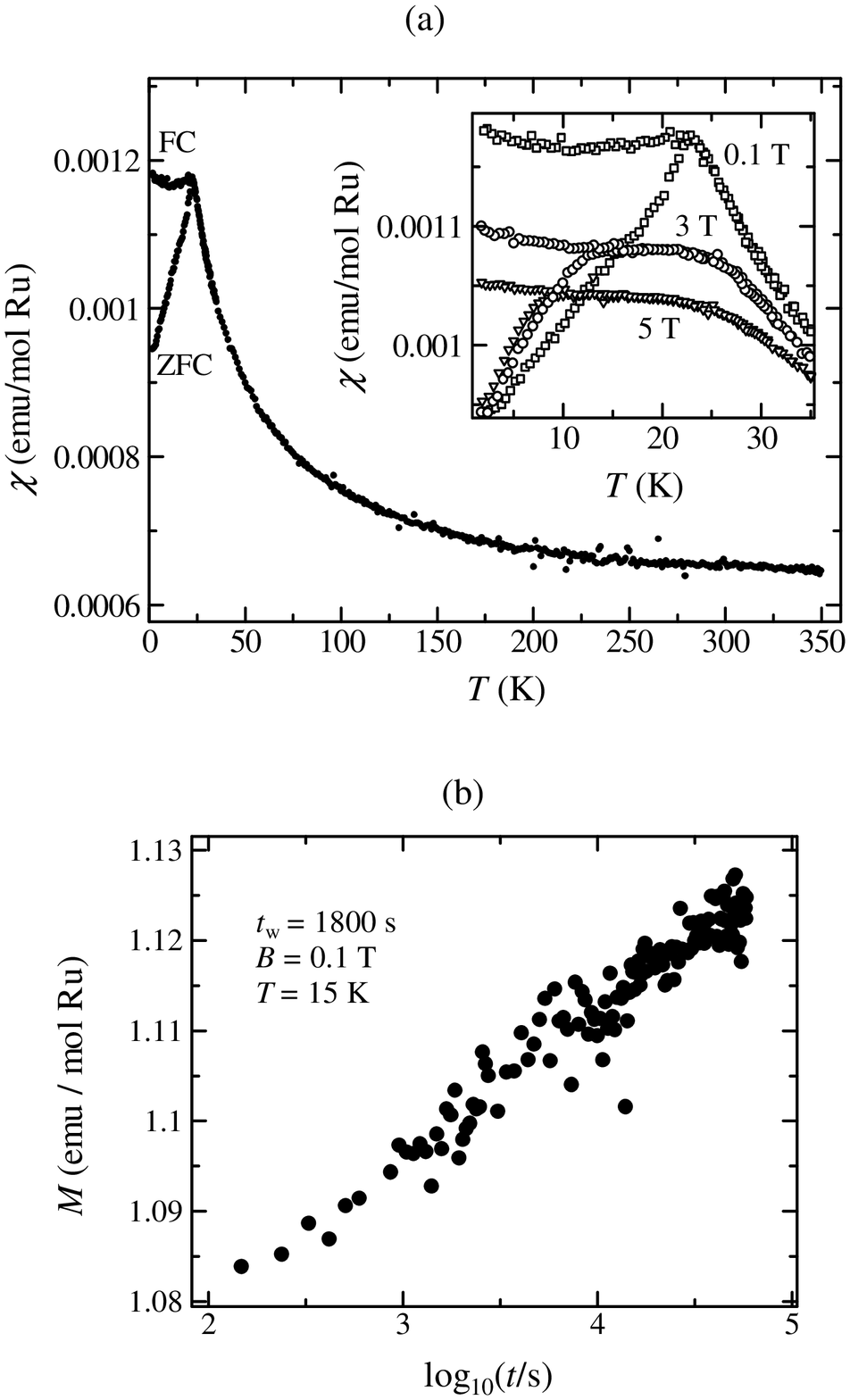}
\end{center}
\caption{(a) Magnetic susceptibility of Ca$_{2}$Ru$_{2}$O$_{7}$ at 0.1 T. The magnetization becomes irreversible below 23 K. The inset shows a comparison of the low-temperature susceptibilities under several magnetic fields. (b) Time dependence of magnetization. The sample was cooled under the ZFC condition down to 15 K. After waiting for 1800 s, a magnetic field of 0.1 T was applied. Then, the magnetization was recorded as a function of time.}
\label{fig2}
\end{figure}

Figure~\ref{fig2}(a) shows the temperature dependence of the magnetic susceptibility.
The susceptibility above 30 K is almost perfectly reproduced using the function
\begin{equation}
\label{eq1}
\chi=\frac{C}{T-\Theta} + \chi_{0},
\end{equation}
where $C$, $\Theta$ and $\chi_{0}$ are the Curie constant, the Weiss temperature and a constant term independent of temperature, respectively. The best-fitted values are $C = 1.65 \times 10^{-2}$ emu$\cdot$K/mol Ru, $\Theta = -4.3$ K and $\chi_{0} = 5.95 \times 10^{-4} $ emu/mol Ru. Assuming a simple ionic model, the oxidation number of ruthenium is 5+. Therefore, there are three 4$d$ electrons per Ru atom, and each Ru atom has a $S=3/2$ spin. In this case, we can expect $\mu_{\mbox{\scriptsize eff}}= 3.87$ $\mu_{\mbox{\scriptsize B}}$. However, the observed $\mu_{\mbox{\scriptsize eff}}$ is only 0.36 $\mu_{\mbox{\scriptsize B}}$, smaller by one order of magnitude. On the other hand, the value of $\chi_{0}$ is comparable to that of typical Pauli paramagnetism in highly correlated oxides, which is consistent with the resistivity result, as shown below. 

The most striking feature is the appearance of a sharp transition with an irreversibility between the field-cooled (FC) curve and the zero-field-cooled (ZFC) curve below 23 K. The ZFC curve exhibits a sharp cusp and decreases with decreasing temperature. On the other hand, the FC curve below 23 K is almost constant. This is a typical behavior of a spin glass. Under a higher magnetic field, the cusp becomes less sharp, and the onset temperature of the irreversibility becomes lower as shown in the inset. The time dependence of the magnetization below the glass-transition temperature is shown in Fig.~\ref{fig2}(b). The magnetization exhibits an aging phenomenon, clearly demonstrating another characteristic of a spin-glass state.

\begin{figure}[tb]
\begin{center}
\includegraphics[width=1.0\linewidth]{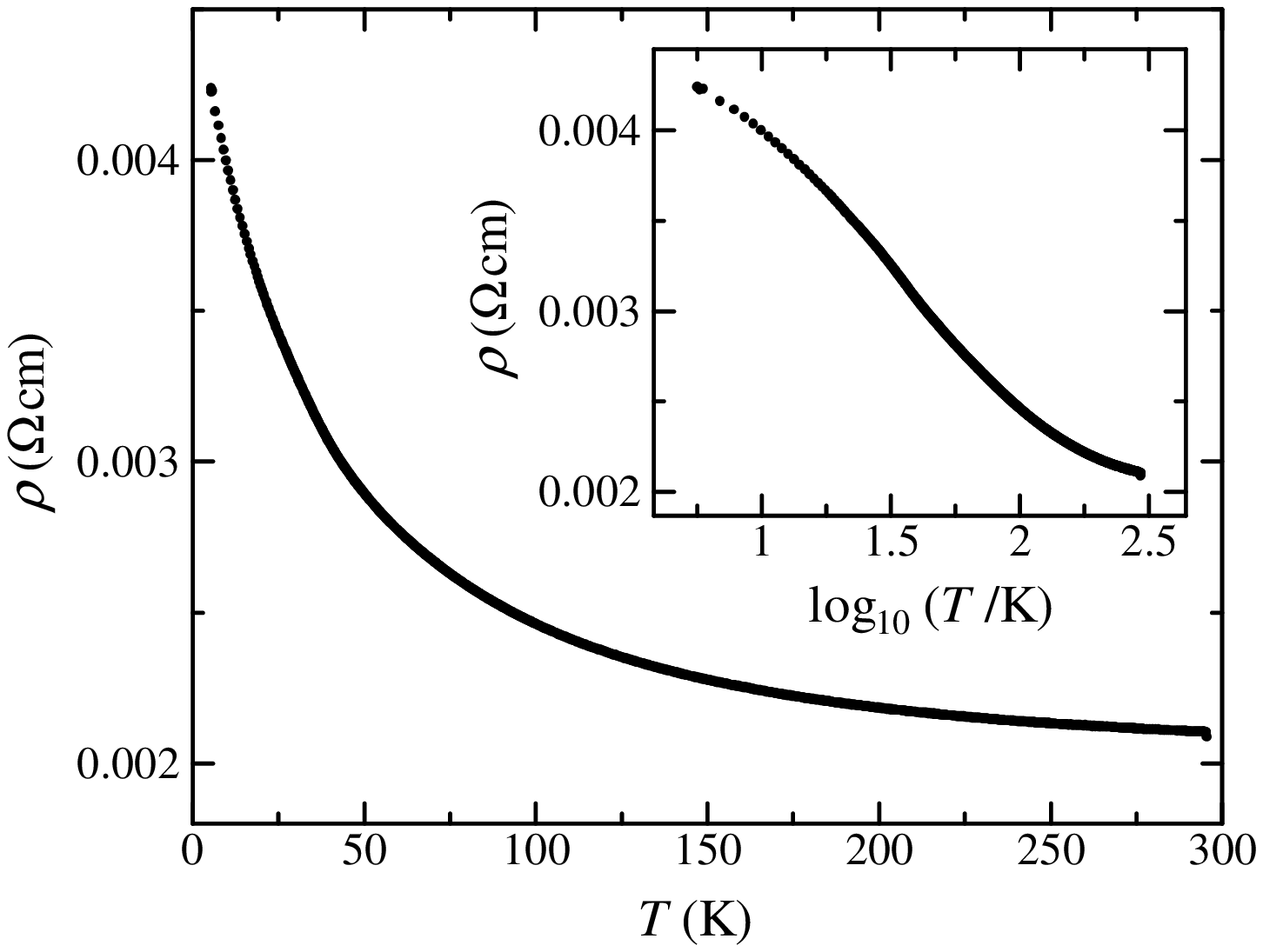}
\end{center}
\caption{Temperature dependence of resistivity of Ca$_{2}$Ru$_{2}$O$_{7}$ single crystal. The resistivity is also plotted as a function of $\log T$ in the inset.}
\label{fig3}
\end{figure}

 Figure~\ref{fig3} shows the temperature dependence of the resistivity of Ca$_{2}$Ru$_{2}$O$_{7}$. The resistivity at room temperature is 2$\times$$10^{-3}$ $\Omega$cm, as large as a typical value for metallic, highly correlated oxides. 
The behavior is not that of a simple metal, because the temperature coefficient is negative over the whole temperature range measured between 5.5 and 300 K. In the lowest temperature region, the resistivity does not appear to saturate but continues to increase with decreasing temperature. However, the rise in the resistivity at low temperatures is less steep; the ratio, $\rho_{\mbox{\scriptsize 5.5K}} / \rho_{\mbox{\scriptsize 295K}}$, is only about 2. Clearly, an activation function, $\rho \propto \exp (E_{\mbox{\scriptsize A}}/k_{\mbox{\scriptsize B}}T)$, does not reproduce the observation, indicating that no gap is formed at the Fermi level. Therefore, the electronic state is basically metallic, and the $\chi_{0}$ term in the susceptibility can be interpreted in terms of Pauli paramagnetism. We also attempted to fit the data with a variable-range hopping funcition, $\rho \propto \exp \{ (E_{\mbox{\scriptsize A}}/k_{\mbox{\scriptsize B}}T)^{(1/(d+1))} \}$, but were unsuccessful. On the other hand, it seems that a $\rho$ vs $\log T$ plot begins to saturate at low temperatures, as shown in the inset. The resistivity shows no anomaly at the glass-transition temperature, $T_{g}$. This is not unusual for a conductive spin-glass system.\cite{taniguchi04}

It is known that Y$_{2}$Ru$_{2}$O$_{7}$ is an insulator but becomes metallic when Bi is substituted for Y.\cite{yoshii99} This is explained in terms of a Mott transition. The Ru-O-Ru bond angles are 129$^{\circ}$ and 139$^{\circ}$ in Y$_{2}$Ru$_{2}$O$_{7}$ and Bi$_{2}$Ru$_{2}$O$_{7}$, respectively.\cite{ishii00} The larger angle gives a broader band width, resulting in a metallic state. Ca substitution was also attempted by the same author. However, the upper limit of $x$ was 0.6 in Y$_{2-x}$Ca$_{x}$$_{2}$Ru$_{2}$O$_{7}$, and the electronic state was still that of an insulator. In Ca$_{2}$Ru$_{2}$O$_{7}$, the Ru-O-Ru bond angle is 135.72$^{\circ}$, larger than that in Y$_{2}$Ru$_{2}$O$_{7}$. This may be one of the reasons why Ca$_{2}$Ru$_{2}$O$_{7}$ is metallic. Furthermore, the change in the band filling is also important. Note that the isovalent Cd$_{2}$Ru$_{2}$O$_{7}$ is also metallic.\cite{wang98}

We now explain the origin of the localized spins, all of which undergo the spin-glass transition. One might interpret the data to conclude that Ca$_{2}$Ru$_{2}$O$_{7}$ itself exhibits only Pauli paramagnetism, and that the coexistence of a small amount of another phase contributes to the spin-glass behavior. However, our sample is an ensemble of small single crystals with well-defined shapes, and the powder X-ray diffraction demonstrated that there is no contamination, as shown in Fig.~\ref{fig1}, except for a very small amount of RuO$_{2}$, which is known to exhibit Pauli paramagnetism.\cite{ryden70} We measured the magnetic susceptibility for three different batches. All the results quantitatively agreed; the ratio of the Curie-Weiss contribution and the Pauli paramagnetism is always the same. If the Curie-Weiss contribution were from an extrinsic origin, this ratio would vary from batch to batch. Therefore, it is most likely that the spin-glass behavior is an intrinsic feature of our phase.

Next, we clarify whether our pyrochlore includes some magnetic impurity atoms in the crystal lattice, because an inclusion of only 0.4 \% of Fe$^{3+}$ ions may cause the observed magnetic moments. We can exclude this possibility because we used reagents of more than 99.99 \% purity. In the Ru powder used in this study, the content of magnetic elements such as Fe was less than 14 ppm. In the CaO powder, it was less than 1 ppm. The H$_{2}$O$_{2}$ solution used was almost free of magnetic impurities.

Whatever the origin of the magnetic moments, it is unlikely that only nearest-neighbor interactions cause the spin-glass state, because the glass temperature is as high as 23 K despite the very low spin concentration. Therefore, there should be Ruderman-Kittel-Kasuya-Yoshida interactions mediated by the conduction electrons. In such a mechanism, the exchange interactions can be ferromagnetic or antiferromagnetic. The coexistence of these interactions reduces the absolute value of the Weiss temperature. In our crystal, the factor $T_{\mbox{\scriptsize g}} / |\Theta|$ is as large as 5.4. This is in contrast with the insulating spin-glass pyrochlore, Y$_{2}$Mo$_{2}$O$_{7}$, in which $T_{\mbox{\scriptsize g}} / |\Theta|$ is as small as 0.11, showing that geometric frustration is predominant in this molybdate.

Although our X-ray structural analysis did not detect any clear evidence, we cannot yet exclude the possibility of a deviation from ideal Ca$_{2}$Ru$_{2}$O$_{7}$ composition. If there are oxygen defects, for example, some of the Ru atoms are deoxidized into Ru$^{4+}$, possessing an excess electron. If these excess electrons behave as localized moments, they can cause a Curie-Weiss magnetism in addition to the Pauli paramagnetism coming from the conduction electrons. Another possibility is that the conduction electrons themselves have weak magnetic polarizations delocalized over many atoms, similar to those in stoner glass\cite{hertz79} and spin-density glass.\cite{sachdev95} In any case, there remains a fundamental question: why did the Ru 4$d$ electrons become partially magnetic despite the metallic conduction? Note that many perovskite-structure metallic ruthenates exhibit anomolous magnetism instead of ordinary Pauli paramagnetism. For example, CaRuO$_{3}$, which has metallic conduction, exhibits Curie-Weiss magnetic susceptibility with weak irreversibility.\cite{felner00} Therefore, the metallic state in Ca$_{2}$Ru$_{2}$O$_{7}$ might be so fragile that a small purturbation causes magnetic moments. This is in contrast with the study on Bi$_{2-x}$$M_{x}$Ru$_{2}$O$_{7}$ ($M=$ Mn, Fe, Co, Ni and Cu), where the substitution with magnetic atoms $M$ does not affect the Pauli paramagnetic state of the Ru-O sublattice of Bi$_{2}$Ru$_{2}$O$_{7}$.\cite{haas02}

 In conclusion, we have discovered a novel pyrochlore ruthenate, Ca$_{2}$Ru$_{2}$O$_{7}$. The magnetic susceptibility exhibits the behavior of a spin glass in addition to Pauli paramagnetism, although the effective magnetic moment of the spin glass was smaller than expected by one order of magnitude. The resistivity of $2 \times 10^{-3}$ $\Omega$cm at room temperature suggests a metallic electronic state, but the $\rho$ vs $T$ curve has negative slope over the whole temperature region, indicating that the conduction electrons are strongly scattered by the magnetic moments. In future studies, we intended to first verify whether the spin-glass behavior is intrinsic in Ca$_{2}$Ru$_{2}$O$_{7}$. Detailed measurements of the transport properties under a magnetic field will be useful, because they may detect chirality in the spin-glass state.\cite{taniguchi04} NMR or neutron-scattering measurements are necessary to clarify the origin of the magnetic moments and the mechanism of the spin glass. Investigation of the electronic phase diagram by doping with Y will also be important.

This study was supported by Grants-in-Aid for Scientific Research No. 15750127 and No. 17540341 from the Ministry of Education, Culture, Sports, Science and Technology.


\begin{thebibliography}{99} 

\bibitem{gingras97} M. J. P. Gingras, C. V. Stager, N. P. Raju, B. D. Gaulin and J. E. Greedan: Phys. Rev. Lett. \textbf{78} (1997) 947.
\bibitem{gardner99} J. S. Gardner, B. D. Gaulin, S. -H. Lee, C. Broholm, N. P. Raju and J. E. Greedan: Phys. Rev. Lett. \textbf{83} (1999) 211.
\bibitem{miyoshi00} K. Miyoshi, Y. Nishimura, K. Honda, K. Fujiwara and J. Takeuchi: J. Phys. Soc. Jpn. \textbf{69} (2000) 3517.

\bibitem{harris97} M. J. Harris, S. T. Bramwell, D. F. McMorrow, T. Zeiske and K. W. Godfrey: Phys. Rev. Lett. \textbf{79} (1997) 2554.
\bibitem{ramirez99} A. P. Ramirez, A. Hayashi, R. J. Cava, R. Siddharthan and B. S. Shastry: Nature \textbf{399} (1999) 333.
\bibitem{higashinaka05} R. Higashinaka and Y. Maeno: Phys. Rev. Lett. \textbf{95} (2005) 237208.


\bibitem{taguchi03} Y. Taguchi, T. Sasaki, S. Awaji, Y. Iwasa, T. Tayama, T. Sakakibara, S. Iguchi, T. Ito and Y. Tokura: Phys. Rev. Lett. \textbf{90} (2003) 257202.
\bibitem{onoda03} S. Onoda and N. Nagaosa: Phys. Rev. Lett. \textbf{90} (2003) 196602.

\bibitem{shimakawa96} Y. Shimakawa, Y. Kubo and T. Manako: Nature \textbf{379} (1996) 53.

\bibitem{sakai01} H. Sakai, K. Yoshimura, H. Ohno, H. Kato, S. Kambe, R. E. Walstedt, T. D. Matsuda and Y. Haga: J. Phys: Cond. Mat. \textbf{13} (2001) L785.
\bibitem{hanawa01} M. Hanawa, Y. Muraoka, T. Tayama, T. Sakakibara, J. Yamaura, Z. Hiroi: Phys. Rev. Lett. \textbf{87} (2001) 187001.

\bibitem{yonezawa04} S. Yonezawa, Y. Muraoka, Y. Matsushita and Z. Hiroi: J. Phys. Soc. Jpn. \textbf{73} (2004) 819.
\bibitem{yonezawa04b} S. Yonezawa, Y. Muraoka and Z. Hiroi: J. Phys. Soc. Jpn. \textbf{73} (2004) 1655.
\bibitem{hiroi04} Z. Hiroi, S. Yonezawa and Y. Muraoka: J. Phys. Soc. Jpn. \textbf{73} (2004) 1651.
\bibitem{hiroi05} Z. Hiroi, S. Yonezawa and Y. Muraoka: J. Phys. Soc. Jpn. \textbf{74} (2005) 3399.

\bibitem{maeno94} Y. Maeno, H. Hashimoto, K. Yoshida, S. Nishizaki, T. Fujita, J. G. Bednorz and F. Lichtenberg: Nature \textbf{372} (1994) 532.
\bibitem{ishida98} K. Ishida, H. Mukuda, Y. Kitaoka, K. Asayama, Z. Q. Mao, Y. Mori and Y. Maeno: Nature \textbf{396} (1998) 658.

\bibitem{longo69} J. M. Longo, P. M. Raccah and J. B. Goodenough: Mater. Res. Bull. \textbf{4} (1969) 191.
\bibitem{cox83} P. A. Cox, R. G. Egdell, J. B. Goodenough, A. Hamnett and C. C. Naish: J. Phys. C \textbf{16} (1983) 6221.
\bibitem{hsu88} W. Y. Hsu, R. V. Kasowski, T. Miller and T. -C. Chiang: Appl. Phys. Lett. \textbf{52} (1988) 792.

\bibitem{aleonard62} R. Al\'{e}onard, E. F. Bertaut, M. C. Montmory and R. Pauthenet: J. Appl. Phys. \textbf{33} (1962) 1205.
\bibitem{subramanian83} M. A. Subramanian, G. Aravamudan and G. V. Subba Rao: Prog. Solid State Chem. \textbf{15} (1983) 55.

\bibitem{yoshii99} S. Yoshii and M. Sato: J. Phys. Soc. Jpn. \textbf{68} (1999) 3034.
\bibitem{ito00} M. Ito, Y. Yasui, M. Kanada, H. Harashina, S. Yoshii, K. Murata, M. Sato, H. Okumura and K. Kakurai: J. Phys. Soc. Jpn. \textbf{69} (2000) 888.

\bibitem{takeda98} T. Takeda, M. Nagata, H. Kobayashi, R. Kanno, Y. Kawamoto, M. Takano, T. Kamiyama, F. Izumi and A. W. Sleight: J. Solid State Chem. \textbf{140} (1998) 182.

\bibitem{wang98} R. Wang and A. W. Sleight: Mater. Res. Bull. \textbf{33} (1998) 1005.

\bibitem{kennedy95} B. J. Kennedy: Acta Cryst. C \textbf{51} (1995) 790.

\bibitem{taniguchi04} T. Taniguchi, K. Yamanaka, H. Sumioka, T. Yamazaki, Y. Tabata and S. Kawarazaki: Phys. Rev. Lett. \textbf{93} (2004) 246605.

\bibitem{ishii00} F. Ishii and T. Oguchi: J. Phys. Soc. Jpn. \textbf{69} (2000) 526.

\bibitem{ryden70} W. D. Ryden and A. W. Lawson: J. Chem. Phys. \textbf{52} (1970) 6058.

\bibitem{hertz79} J. A. Hertz: Phys. Rev. B \textbf{19} (1979) 4796.

\bibitem{sachdev95} S. Sachdev, N. Read and R. Oppermann: Phys. Rev. B \textbf{52} (1995) 10286.

\bibitem{felner00} I. Felner, I. Nowik, I. Bradaric and M. Gospodinov: Phys. Rev. B \textbf{62} (2000) 11332.

\bibitem{haas02} M. K. Haas, R. J. Cava, M. Avdeev and J. D. Jorgensen: Phys. Rev. B \textbf{66} (2002) 094429.

\end{thebibliography}
\end{document}